# U.S. academic libraries: understanding their web presence and their relationship with economic indicators

**Enrique Orduña-Malea[1*] and John J. Regazzi[2]**

[1]Department of Audiovisual Communication, Documentation, and History of Art. Polytechnic University of Valencia (UPV), Valencia, Spain.
Camino de Vera s/n, Valencia 46022, Spain.
* e-mail: enorma@upv.es

[2]Palmer School of Library and Information Science and Department of Computer Science and Management Engineering. Long Island University, 520 Northern Blvd., Brookville, NY 11552, USA.

**Abstract**
The main goal of this research is to analyze the web structure and performance of units and services belonging to U.S. academic libraries in order to check their suitability for webometric studies. Our objectives include studying their possible correlation with economic data and assessing their use for complementary evaluation purposes. We conducted a survey of library homepages, institutional repositories, digital collections, and online catalogs (a total of 374 URLs) belonging to the 100 U.S. universities with the highest total expenditures in academic libraries according to data provided by the National Center for Education Statistics (NCES). Several data points were taken and analyzed, including web variables (page count, external links, and visits) and economic variables (total expenditures, expenditures on printed and electronic books, and physical visits). The results indicate that the variety of URL syntaxes is wide, diverse and complex, which produces a misrepresentation of academic libraries' web resources and reduces the accuracy of web analysis. On the other hand, institutional and web data indicators are not highly correlated. Better results are obtained by correlating total library expenditures with URL mentions measured by Google (r= 0.546) and visits measured by Compete (r= 0.573), respectively. Because correlation values obtained are not highly significant, we estimate such correlations will increase if users can avoid linkage problems (due to the complexity of URLs) and gain direct access to log files (for more accurate data about visits).

## Introduction

A new university ranking, the "Webometrics Ranking of World Universities", was published by the Cybermetrics Lab in 2004, and described by Aguillo et al. (2008). This ranking, based on webometric methods, considers documentation published and accessible via the web, and specifically the size (page count) and impact (external inlinks) of such resources. All institutions belonging to a university –such as faculties, schools, and research groups– contribute with content and impact to the general web presence of the university. In this context, the university can be understood as a complex online system that may be measured through its official website.

The contribution of academic libraries to the web performance of the corresponding university is estimated *a priori* to be very high, due to the large amount of content stored on the library website. The digitalization of printed content and the creation of digital collections conform to a wide set of digital issues. Indeed, within the "Code of best practices in fair use for academic and research libraries," published by the ARL[1], one of the practices described clearly calls for "collecting material posted on the World Wide Web and making it available." Moreover, Regazzi (2012a; 2012b) has recently pointed out the increase in electronic expenditures in academic libraries, which may result in more online content being detected and measured by webometric techniques.

In this sense, the webometric analysis of academic libraries, along with other purposes, can be used for the following activities:
- To describe and measure the library's website performance. Moreover, if web data correlated with economic and institutional data (such as expenditures, holdings, etc.), a relationship between economic investment and web performance could be established.
- To analyze online information to detect areas of the institution that may be measured through webometric techniques, such as financial information (Gallego et. al 2009).
- To determine the degree to which the academic library is contributing to the general web performance of the university (and therefore, their position in web rankings).

This paper focuses on the first point, library website performance, leaving the remaining topics for future research.

**Review of the Literature**

Webometric methodologies, despite having been widely applied to international academic environments, are seldom used within the U.S. university system. Webometric methods are widely applied in Europe (Ortega and Aguillo 2009a; Thelwall and Zuccala 2008), and to a lesser extent in other continents such as Africa (Adecannby 2011), Asia (Qiu et. al 2004), and Australia (Smith and Thelwall 2002). In the United States and Canada, Ortega and Aguillo (2009b) have carried out notable work in analyzing US universities. Particular university units such as departments of librarianship and information science have been explored by Arakaki and Willet (2009), and Chu and Thelwall (2002).

If we focus on academic libraries, we can observe that several researchers have studied the information architecture, the Web 2.0 phenomenon, and web usability within academic library websites (Harinarayana and Raju 2010; Alton et. al 2010; Mahmood and Richardson 2012). All these issues are of interest to webometrics because an increase in the number of social services indirectly affects the number of user interactions (reflected by an increased number of total visits to the website, mentions, and links). These topics have also been studied in other types of libraries, such as national libraries (Buigues-Garcia and Gimenez-Chornet 2012).

Nonetheless, research on U.S. academic libraries from a webometric perspective has received limited attention. Tang and Thelwall (2008) conducted pioneer research analyzing the patterns of links from and to the websites of 100 U.S. academic and public libraries finding on one hand a significant relation between visibility and page count indicators for academic libraries, and on the other hand little interaction between U.S. universities and public libraries.

Moreover, other units emerge within academic library websites, suitable for storing and making available large amounts of digital contents, mainly online catalogs, digital collections and repositories. Due to the increasing implantation of institutional repositories to make the research output of universities available, a special interest in this unit is detected in the literature in comparison to the other units (in webometric terms), especially for web usage measurements. In this regard, Scholze (2007) pointed out some methods in which usage data for repositories could be collected, and Zuccala et. al (2007) used web link analysis and log files to study the impact and usage of an institutional repository (the National electronic Library for Health).

The project of the Ranking Web of Repositories[2], launched by the Cybermetrics Lab in Spain, should also be mentioned (Aguillo et. al 2010). This project ranks institutional and thematic repositories according to the quantity of files stored (especially PDF files), presence on Google Scholar and number of external links received. One of the main findings of these authors is the inability of search engine crawlers to collect data due to barriers in the design of the web databases. This problem is also studied by Arlitsch and O'Brian (2012), who indicate that the usage of non-recommended metadata schemes is the cause of the invisibility of repositories in Google Scholar.

Other studies have focused on the analysis of the origin of links to institutional repositories (Smith 2012). Sato and Itsumura (2011) analyze links to the Kyoto University institutional repository, finding that links were fundamentally made from non-academic sources such as Wikipedia and personal web pages. Smith (2011) also found that a significant number of links to institutional repositories were made from non-research sites such as Wikipedia. This influence of Wikipedia is aligned with the findings of Orduña-Malea and Ontalba-Ruipérez (2012), which show that links from Wikipedia to Spanish universities could be used as substitutes of total external links.

If we focus on the web performance analysis of U.S. academic libraries (and their internal units), the following research gaps -not previously treated- are detected: a) the suitability of academic library websites for webometric analysis (by means of their URL structures), and b) the correlation between web data and economic variables. If such a correlation is identified, another study might detect and predict online performance.

**Objectives**

The main goal of this research is to analyze the web structure and performance of units and services belonging to U.S. academic libraries (by means of their URLs) in order to estimate their suitability for webometric studies, and their possible correlation with economic and institutional data.

The specific objectives are set out below:
- To identify and describe the variety of URL syntaxes of different academic library web units studied.
- To analyze the presence and distribution of different typologies of academic library web units (library homepages, repositories, OPACs, and digital collections), and their web performance (by page count, mention, impact, and usage indicators, described in the method section), measured through the URLs collected previously.
- To study the correlation among web indicators to detect their coherence and/or differences in describing the web performance of the library web units analyzed.
- To study the correlation among economic indicators (by spending, holdings, and physical visits indicators) to detect their coherence and/or differences in describing the performance of the library web units analyzed, paying special attention to the

- expenditures on electronic material and their possible influence on the web performance of these institutions.
- To detect the relationship – if any – between web data of U.S. academic libraries and economic data.

**Methodology**

The paper analyzes the resources of U.S. academic libraries, taking into account two types of data: institutional and economic data, and web-based data.

The methodology is divided into three main sections. First, we describe the procedure followed to obtain the sample of academic libraries (and their corresponding URLs). Second, we show the indicators used to measure and the procedure used to apply them to each URL. Finally we demonstrate the statistical analysis carried out with the data obtained in the previous step.

*Obtaining the sample*

A sample of 100 universities was selected, considering the total expenditures in academic libraries in 2008 (last available data). This data was extracted from the survey "Documentation for the academic libraries survey: Public use data file / Fiscal year 2008" (Phan et. al 2009). All universities considered in that report were sorted by "total expenditure variable (TE)," and only the first 100 were selected.

Once the 100 universities had been selected, their corresponding URLs were listed and a second step was taken, which consisted of searching the academic library websites within each of the listed universities.

Other services and subunits (suited to storing large amounts of online documents) are commonly created under different domains within the academic library website. For that reason, we established that the term "academic libraries" also considers the following services/products/institutions:
- Library (general, branches, or for specific schools or faculties).
- OPAC or online catalog and searcher.
- Institutional repositories.
- Digital collections.

The identification of each of these institutions (and their corresponding URLs) was achieved by browsing and searching the websites of each of the previously listed 100 universities. This process was carried out in December 2011.

Only units with a sub-domain within the academic website (<xxx.academicdomain.edu>) were considered; this limitation was necessary in order to proceed with an accurate link analysis.

Each university and library unit presented specific considerations in the creation of the websites, providing a wide variety of URL syntaxes which have a direct implication on the content structure and hierarchy of units and institutions.

Next, all types of URL syntaxes detected per unit are identified, classified, and exemplified, explaining and justifying the procedure followed in each case (URL included or eliminated), always taking into account the selection criterion previously commented on (independent sub-domain within the university website). This procedure aims to fulfill the first specific objective.

*Libraries*

Only sub-domains of general academic websites were considered; for this reason, all units outside it were eliminated. For example:

> *University of Louisiana*
> <louisianadigitallibrary.org> (URL outside the university website).

Subdirectories were also eliminated. For example:

> *Baylor University*
> <baylor.edu/**lib**>
>
> *Nova Southeastern University*
> <nova.edu/**library**>

In some cases, redirections from sub-domains to sub-folds (or vice versa) were detected:

> *University of Louisville*
> <**library**.louisville.edu> redirects to:
> <louisville.edu/**library**>
>
> *East Carolina University*
> <**lib**.ecu.edu> redirects to:
> <ecu.edu/**lib**>

In these cases, only sub-domains were kept, and sub-folds were eliminated.

For those universities with more than one active URL (multi-domain activity), we expected to find a corresponding active library URL. Notwithstanding, we detected that some libraries had only activated the corresponding website within one of the existing university web domains. For that reason, a manual check for each university URL version was performed in order to verify the existence of the corresponding library URL version. For example:

> *Southern Illinois University at Carbondale*
> URL 1: <siuc.edu>
> URL 2: <siu.edu>
> Library URL: <lib.siu.edu> (any <lib.**siuc.edu**> exists).

In some universities, library branches, OPACs, digital collections, and repositories were created as sub-domains of the academic library. For example:

> *Oklahoma State University*
> University URL: <okstate.edu>
>   Library URL: <**library**.okstate.edu>
>     Repository URL: <**e-archive**.library.okstate.edu>
>     OPAC URL: <**osucatalog**.library.okstate.edu>

In these cases, only the academic library URL was considered (third level), because other units were contained within it.

Some libraries (and library services) were located within the School to which they belong, such as Business and Law Schools, which typically had their own sub-domain. Some cases of sub-folds were detected, which were eliminated. For example:

> *University of Missouri*
> <law.missouri.edu/**library**>
>
> *Southeastern University*
> <nsulaw.nova.edu/**library**>

In other cases, a sub-domain was identified (which was also considered valid in this study), if it occurred either below or above the School sub-domain. For example:

> *University of Alabama*
> <**library.law**.ua.edu>
>
> *SUNY at Buffalo*
> <**law.lib**.buffalo.edu>

The same considerations were taken into account for OPACs:

> *Howard University*
> <**daniel.law**.howard.edu>

and for repositories:

> *Texas Tech University*
> <**repository.law**.ttu.edu>

Moreover, services may remain outside the School. For example:

> *Fordham University*
> <**lawpac**.fordham.edu>

*OPACs and online catalogs*

With regard to online catalogs, there was also great variety:

a) The URL of the OPAC could be a first-level sub-domain of the academic library (second-level for the university). In this case, only the general URL of the library was considered. For example:

*University of Miami*
<**catalog.library**.miami.edu>

*Brigham Young University*
<**catalog.lib**.byu.edu>

Within this category, another variety was detected. The catalog could be expressed as a second-level domain of the University, but the first-level might not be accessible. For example:

*SUNY at Buffalo*
<**lib**.buffalo.edu> (without direct access to any content).
<**catalog.lib**.buffalo.edu> (valid URL).
<**law.lib**.buffalo.edu> (valid URL).

*University of Utah*
<**library**.utah.edu> (without direct access to any content).
<**search.library**.utah.edu> (valid URL).
<**libtools.library**.utah.edu> (valid URL).
<**libraryfind.library**.utah.edu> (valid URL).

b) The URL of OPAC was independent of the library (outside the library website). For example:

*University of California - Riverside*
<**scotty.ucr**.edu>

*University of Rice*
<**cordoba.rice**.edu>

c) The URL was outside the Library and the University, and for that reason, it was not considered. For example:

*Louisiana State University*
<lsu.**louislibraries.org**>

*Boston University*
<buprimo.hosted.**exlibrisgroup.com**>

d) Some universities maintained the existence of "alias" domains. For example:

*University of Rochester*
<**library**.rochester.edu>
<**lib**.rochester.edu>

And their corresponding subdomains for the online catalog were working, and valid:

<**catalog**.library.rochester.edu>
<**catalog**.lib.rochester.edu>

*Institutional repositories*

Institutional repositories could be observed with more or less the same variety as for online catalogs. Again, several examples should be considered, as shown below.

a) First-level subdomains of the academic library were not considered (as it was for online catalogs, previously shown). For example:

*Duke University*
<**dukespace.lib**.duke.edu>

*Johns Hopkins University*
<**jscholarship.library**.jhu.edu>

As for OPACs, only a general academic library web domain was considered in these cases.

b) The URL of the institutional repository was independent of the library. For example:

*Harvard University*
<**dash**.harvard.edu>

*Columbia University*
<**academiccommons**.columbia.edu>

c) The URL of the repository was outside both the library and the university, and for that reason, it was not considered. For example:

*Fordham University*
<fordham.**bepress.com**>

*University of Houston*
<repositories.**tdl.org/uh-ir**> (which belongs to *Texas Digital Library*).

d) Some institutions maintained the existence of alias domains. For example:

*Rice University*
<**dspace**.rice.edu> and <**rudr**.rice.edu> redirected to:

    <**scholarship**.rice.edu>

In this case, all three sub-domains were valid and taken into account.

*George Mason University*
<**mars**.gmu.edu> redirected to:
<**digilib**.gmu.edu>

e) Repositories for university systems:

Some universities with branch campuses might assign an URL to each campus, but might still share the same repository in a different URL. For example:

System: *Indiana University*
<iu.edu>
Campus: *Indiana University – Bloomington*
<iub.edu>
Repository: *IU Scholar Works*
<scholarworks.**iu.edu**>

In this case, this repository could not be associated with Bloomington, because the URL domain was different.

Other examples of shared repositories, not taken into account, include:

*University of California*
<escholarship.org>

*SUNY*
<dspace.sunyconnect.suny.edu>

*Digital collections*

Digital collections appeared generally in two different ways:

a) First-level sub-domain of the academic library:

    *Yale University*
    <**digitalcollections.library**.yale.edu>

    *UCLA*
    <**digital2.library**.ucla.edu>

b) The URL of the repository was independent of the library. For example:

   Digital scriptorium (*Columbia University*)
   <**scriptorium**.columbia.edu>

Sometimes, despite the existence of a web domain to gather all digital collections, a URL could be found to house independent collections. For example:

*Harvard University*
General: <**digitalcollections**.harvard.edu>
Independent collection: <**arboretum**.harvard.edu>

*Measuring the sample*

This section describes the indicators and sources used to measure the library units collected previously. First the institutional and economic data is introduced, and then the web-based indicators.

*Institutional and economic data*

From the survey "Documentation for the academic libraries survey (ALS): Public use data file. Fiscal Year 2008" (Phan et. al 2009), conducted by the Institute of Education Science (IES), we extracted the sample of 100 universities and sorted them by total expenditure variables (TE).
 Additionally, we recovered, for each university's academic libraries, the following parameters (Phan et. al 2009):
- Expenditures on books.
- Expenditures on electronic books and other electronic materials (one-time purchases).
- Expenditures on current serial subscriptions (ongoing commitments).
- Expenditures on electronic serials.
- Physical visits in a typical week: report the number of persons who physically enter library facilities in a typical week. It is understood that a single person may be counted more than once.

*Web data*

Table 1 shows the different web indicators used (grouped into categories), the sources to obtain them, and the commands employed to query each source.

Table 1. Summary of categories, indicators, sources, and commands used for web measures

| CATEGORY | PAGE COUNT | | MENTION | | IMPACT | USAGE |
|---|---|---|---|---|---|---|
| **INDICATOR** | **TOTAL** | **PDF** | **URL MENTION** | **INLINKS** | **DmR** | **WEB VISITS** |
| **SOURCE** | *Yahoo* | *Google* \| *Yahoo* | *Google* \| *Yahoo* | *MAJESTIC* \| *OSE* | *OSE* | *Compete* |
| **COMMAND** | <site:domain.edu> | <site:domain.edu - filetype:pdf> | <"domain.edu" – site:domain.edu> | Direct | Direct | Direct |

The webometric analysis of each URL was carried out manually in December 2011. The explanation of each measurement is shown below:

Size

Size refers to the number of files in the web domain studied. In this case, we considered two specific indicators:

- **Total**: total number of files within a web domain.

*Google*, despite currently being the search engine with the highest coverage in the world, has flaws in retrieving total page count, due to the elimination of pseudo-duplicates (files that seem the same, but are different), and it has a deficiency in working with non-friendly dynamic URLs (Thelwall 2009). For this reason, only *Yahoo!* was used.

- **PDF**: total number of PDF files.

Employed because this format is the main vehicle for publishing final intellectual works, such as books, papers, and so on (Aguillo 2009). In this case, both *Google* and *Yahoo!* were used. *Google* provides more results and it is more reliable. *Yahoo!* was taken into account in order to have comparative data.

Mention

This category refers to the number of times that the object of analysis (in this case, each web unit) is mentioned in any other online file.

In this case, we considered the two main mention indicators: external inlinks and URL mentions.

- **External inlinks**: the number of links that come from external websites.

At this moment, only *Majestic*[3] and *Open Site Explorer*[4] provide this metric (*Ahrefs Site Explorer*[5] still has low coverage). In the past, *Yahoo Site Explorer* was the main tool, but it was disabled in 2011.

- **URL mentions**: the number of times that the URL is mentioned in any online content.

At present, some research is being pursued on the validity of this indicator as a predictor of external links (Thelwall and Sud 2011). Although there are some problems, *Google* and *Yahoo!* provide the best and most reliable results.

Impact

This category refers to the power of each website through a metric which takes into account not only the quantity but also the quality of the external inlinks received. For example, *PageRank* (PR) is such an impact indicator.

In this case, only *Domain MozRank* (DmR) was used, because it is free, the tool has coverage for all institutions analyzed, and the power of discrimination is better than PR (only from 0 to 10).

- **DmR** (*Domain MozRank*)

Offered by *Open Site Explorer* tool, it reflects the importance (from 1 to 100 points) of any given web page on the Internet.

Usage

This category (named also traffic, audience, or popularity) counts the number of visits to a website.

- **Unique visitors**: total number of different users who visited the webpage in a period of time.

In this case, data was extracted from *Compete*[6], due to its good coverage of U.S. institutions, and because it is free. Neither *Alexa*[7] nor other services provided more accurate data.

*Analyzing the sample*

All data gathered were exported into an Excel spreadsheet for analysis. Additionally, an XLStat application was used to carry out the correlation analysis among institutional and web data. A Spearman coefficient was used instead of Pearson since the web data is typically skewed and non-normal.

**Results**

The results obtained were divided into three main sections in order to clarify the information provided: institutional data, web data, and finally, the correlation between them.

*Institutional data*

Harvard University (117,884,296) is the institution with highest total expenditures in academic libraries, followed by Yale (92,247,666), and Stanford (78,376,769). The complete results for all previously selected institutional indicators (total expenditure, expenditure on books and electronic books, expenditure on serials and electronic serials and physical visits) are provided in Annex 1 (available online)[8].

Five universities displayed a lack of data about electronic book expenditures (Stanford University, University of California-Berkeley, University of Georgia, University of Missouri-Columbia, and Fordham University), and two displayed a lack of data about electronic serial expenditures (Harvard and Stanford). We eliminated these institutions from the list, and performed the following correlation analysis (Table 2).

**Table 2. Correlation between institutional data indicators**

|  | Total expenditures | e-Book expenditures | Book expenditures | Serial expenditures | e-Serial expenditures | Physical visits |
|---|---|---|---|---|---|---|
| Total expenditures | 1 | | | | | |
| e-Book expenditures | **0.350\*\*** | 1 | | | | |
| Book expenditures | **0.710\*\*** | **0.457\*\*** | 1 | | | |
| Serial expenditures | **0.775\*\*** | **0.343\*\*** | **0.422\*\*** | 1 | | |
| e-Serial expenditures | **0.644\*\*** | **0.262\*** | **0.416\*\*** | **0.720\*\*** | 1 | |
| Physical visits | **0.389\*\*** | 0.045 | **0.284\*** | **0.240\*** | **0.251\*** | 1 |

\* Significant values (except diagonal) at the level of significance alpha=0.050 (two-tailed test)
\*\* Significant values (except diagonal) at the level of significance alpha=0.001 (two-tailed test)

The results display a generally good correlation between total expenditures and book expenditures (r= 0.710) and serial expenditures (r= 0.775). A good correlation between serial and electronic serial expenditures (r= 0.720) is of particular interest, as is a worse correlation between book and electronic book expenditures (r= 0.475).

A low correlation (the only non significant at alpha= 0.050) was also observed between physical visits and expenditures on electronic books (r= 0.045). This result may indicate a reduction of physical visits to those libraries with greater investment in eBooks (and consequently, a greater collection of available eBooks). In any case, further research is needed in order to verify this hypothesis.

*Web data*

*Distribution of web unit categories*

A total of 374 URLs were retrieved from the 100 selected universities, using the gathering process described in the method section. All web data for all these URLs are available in Annex 2 (available online)[8].

Table 3 shows the percentage of each type of URL considered. A total of 159 URLs belong to academic library homepages, whereas 85 belong to online catalogs. Finally, 65 URLs each were retrieved from digital collections and institutional repositories.

**Table 3. Distribution of the different types of URLs within academic library websites**

| TYPE OF URL | n | % |
|---|---|---|
| Library | 159 | 43 |
| Catalog | 85 | 23 |
| Digital collection | 65 | 17 |
| Repository | 65 | 17 |
| TOTAL | 374 | 100 |

*Missing data*

Table 4 shows the number of URLs without data according to each web indicator. It should be noted that the total missing data value includes cases of both zero matches (the URL is covered, but without results) and no data available (the URL is not covered in the web source). By means of illustration, the latter is shown in parentheses. We also display the percentage according to the global number of URLs analyzed.

**Table 4. Missing data for each web indicator**

| Indicator | Missing data (no data) | % |
|---|---|---|
| Page count (Y) | 10(0) | 2.7 |
| PDF (G) | 142(0) | 38.0 |
| PDF (Y) | 157(0) | 42.0 |
| URL (G) | 0(0) | 0.00 |
| URL (Y) | 6(0) | 1.6 |
| Links (M) | 9(3) | 2.4 |
| Links (O) | 42(36) | 11.2 |
| DmR (O) | 38(36) | 10.2 |
| Visits (C) | 166 (6) | 44.4 |

Three parameters (PDFs, measured by both Google and Yahoo!, and total visits) become the worst indicators. A total of 166 URLs (44.4%) have no data for visits measured by Compete whereas 157 URLs (42%) have no results if measured by number of PDF files in Yahoo, and 142 (38%) in Google.

These results could be explained by the nature of each type of URL considered. For example, OPAC URLs are very often invisible to users; they only appear when a query is performed in the online catalog. For that reason, it is unlikely to be linked or mentioned. Likewise, URLs that belong to digital collections probably do not contain PDF files due to the multimedia or graphic nature of the collections.

Furthermore, the coverage of the source also influences the measurements: Open Site Explorer has missing data from 36 URLs while Compete (6) and MajesticSEO (3) show missing URLs as well.

In order to clarify the influence of the type of URL in the missing data percentage (shown in Table 4), in Table 5 we provide the number of URLs without data for each of the three problematic indicators previously considered.

**Table 5. Distribution of missing data per source and type of URL**

| TYPE | PDF (G) | | PDF (Y) | | Visits (C) | |
|---|---|---|---|---|---|---|
| | No data | % | No data | % | No data | % |
| **Catalog** | 51 | 60.0 | 55 | 64.7 | 42 | 49.4 |
| **Collection** | 29 | 44.6 | 33 | 50.8 | 37 | 56.9 |
| **Library** | 35 | 22.0 | 42 | 26.4 | 46 | 28.9 |
| **Repository** | 27 | 41.5 | 27 | 41.5 | 35 | 53.8 |

Table 5 demonstrates the poor performance of online catalog webpages: 60% of catalog URLs gathered contain no available data for PDFs on Google, and 64.7% on Yahoo. When total visits are considered, digital collections and repositories should be noted for their low results.

*Units' performance*

Table 6 shows the ranking of URLs according to page count (Yahoo!), links (Majestic), and URL mentions (Google).

**Table 6. Top URLs per web indicator: page count, links, and mentions**

| Page count (Y) | | Links (M) | | URL (G) | |
|---|---|---|---|---|---|
| **citeseerx.ist.psu.edu** | 290,000 | **citeseerx.ist.psu.edu** | 3,689,486 | **citeseerx.ist.psu.edu** | 15,300,000 |
| **library.upenn.edu** | 154,000 | **ufdc.ufl.edu** | 112,498 | **library.ucsb.edu** | 13,000,000 |
| **lib.umn.edu** | 149,000 | **dsal.uchicago.edu** | 71,016 | **library.upenn.edu** | 11,800,000 |
| **tvnews.vanderbilt.edu** | 149,000 | **docsouth.unc.edu** | 61,526 | **lib.umich.edu** | 7,980,000 |
| **libraries.mit.edu** | 131,000 | **dl.lib.brown.edu** | 60,501 | **lib.utexas.edu** | 6,600,000 |
| **lib.umich.edu** | 129,000 | **library.duke.edu** | 49,653 | **lib.msu.edu** | 5,540,000 |
| **ufdc.ufl.edu** | 104,000 | **elibrary.unm.edu** | 49,435 | **library.cornell.edu** | 4,190,000 |
| **elibrary.unm.edu** | 102,000 | **texashistory.unt.edu** | 43,167 | **dspace.mit.edu** | 3,560,000 |

| lib.utexas.edu | 87,500 | scriptorium.lib.duke.edu | 33,787 | lib.umn.edu | 3,470,000 |
| lib.virginia.edu | 77,600 | libraries.mit.edu | 32,220 | lib.byu.edu | 3,060,000 |

These results confirm the predominance of CiteseerX, the repository hosted at Pennsylvania State University, which stands at the top of the three indicators displayed. Despite this result, unexpectedly, the remaining repositories appear in low positions, in fact, only three repositories appear in the top 25 if their ranking is considered according to their count (Yahoo) indicator. Apart from CiteseerX, the other repositories are Knowledge Bank (the repository of the Ohio State University, at 16$^{th}$ position), and Dspace@MIT (the repository of the Massachusetts Institute of Technology, at 17$^{th}$ position).

In order to complement this data, the distribution of URLs per type -using Page count (Yahoo!) as ranking criteria- is provided in Table 7. By means of a simple description of the distribution per type, the number of URLs in the top 25, 50, 100, 200 and total URLs is shown.

Table 7. Distribution of URLs per type, using Page count (Yahoo!) as ranking criteria

| TYPE | TOP 25 | TOP 50 | TOP 100 | TOP 200 | TOTAL |
|---|---|---|---|---|---|
| **Catalog** | 4 | 4 | 4 | 9 | 85 |
| **Collection** | 4 | 5 | 9 | 24 | 65 |
| **Library** | 14 | 32 | 69 | 122 | 159 |
| **Repository** | 3 | 9 | 18 | 45 | 65 |

If the top 100 are considered, the number of catalogs' URLs is still the same as the top 25 (four URLs): <novacat.nova.edu>, <vufind.carli.illinois.edu>, <ageconsearch.umn.edu>, and <searchworks.stanford.edu>, proving the low performance of the remaining OPACs, most of which are distributed below position 200. In the case of digital collections, the results are similar: a few URLs present a high count data (four within the top 25), but most of them are in the low positions.

Table 8 shows the same ranking but takes the Link (Majestic) indicator into consideration. The data obtained clearly show an improvement in both catalog and digital collection URLs in the top 50 and 100. Repositories perform practically the same as in Table 7 but have improved their performance in the top 50.

Table 8. Distribution of URLs per type, using Link (Majestic) as ranking criteria

| TYPE | TOP 25 | TOP 50 | TOP 100 | TOP 200 | TOTAL |
|---|---|---|---|---|---|
| **Catalog** | 4 | 13 | 20 | 42 | 85 |
| **Collection** | 5 | 8 | 16 | 31 | 65 |
| **Library** | 13 | 25 | 45 | 83 | 159 |
| **Repository** | 3 | 4 | 19 | 44 | 65 |

*Correlation between web indicators*

On the other hand, Table 9 illustrates the correlation between web indicators, considering only URLs with data in all indicators (up to 325). We have highlighted the low correlation of Count (Y) and URL (Google & Yahoo) with link-based indicators. Moreover, URL mention values correlate highly (r= 0.850) between the different sources used to obtain them (Google and Yahoo), and Count (Y) also shows a high correlation with URL mentions, both for Google (r= 0.836) and Yahoo (r= 0.742).

Table 9. Correlation between web data (n= 325)

|              | Page count (Y) | URL (G)  | URL (Y)  | LINK (M) | LINK (O) | DmR |
|--------------|----------------|----------|----------|----------|----------|-----|
| **Page count (Y)** | 1        |          |          |          |          |     |
| **URL (G)**  | 0.836**        | 1        |          |          |          |     |
| **URL (Y)**  | 0.742**        | 0.850**  | 1        |          |          |     |
| **LINK (M)** | 0.284**        | 0.337**  | 0.301**  | 1        |          |     |
| **LINK (O)** | 0.346**        | 0.370**  | 0.345**  | 0.823**  | 1        |     |
| **DmR**      | 0.281**        | 0.341**  | 0.329**  | 0.741**  | 0.836**  | 1   |

\* Significant values (except diagonal) at the level of significance alpha=0.050 (two-tailed test)
\*\* Significant values (except diagonal) at the level of significance alpha=0.001 (two-tailed test)

*Correlation among institutional and web data*

Finally, this section provides the correlation among institutional and web data for academic libraries. Table 10 shows the correlation, with the "x" axis representing the web indicators, and the "y" axis, the institutional indicators.

For this analysis, the URLs without data in any of the indicators considered have been omitted (91 URLs conform to this analysis), and the performance of different URLs from the same university have been added to obtain a unique value for each university considered.

Table 10. Correlation among institutional and web data (n = 91)

|                        | Page count (Y) | PDF (G)  | PDF (Y)  | URL (G)  | URL (Y)  | LINK (M) | LINK (O) | Web visits (C) |
|------------------------|----------------|----------|----------|----------|----------|----------|----------|----------------|
| Total expenditures     | **0.478**      | **0.375**   | **0.321***  | **0.572**   | **0.472**   | **0.333***  | **0.444**   | **0.570**         |
| e-Book expenditures    | 0.070          | 0.206    | 0.056    | 0.096    | 0.175    | 0.172    | 0.155    | 0.169          |
| Book expenditures      | **0.263***     | **0.371**   | **0.244***  | **0.431**   | **0.420**   | **0.216***  | **0.324***  | **0.444**         |
| Serial expenditures    | **0.321***     | **0.288***  | **0.281***  | **0.390**   | **0.303***  | **0.303***  | **0.328***  | **0.374**         |
| e-Serial expenditures  | **0.227***     | 0.168    | **0.252***  | **0.277***  | 0.200    | 0.166    | **0.218***  | **0.298***        |
| Physical visits        | 0.152          | 0.144    | 0.112    | **0.272***  | **0.307***  | 0.001    | 0.131    | **0.260***        |

\* Significant values (except diagonal) at the level of significance alpha=0.050 (two-tailed test)
\*\* Significant values (except diagonal) at the level of significance alpha=0.001 (two-tailed test)

The results confirm a generally poor correlation. Total expenditures and book and serial expenditures are the economic indicators which achieve significance at α= 0.005 with all web indicators (although the correlations achieved are not very high). On the other hand, URL (G) and Web visits are the web indicators that fit best with economic data. The correlations between total expenditures and both URL (G) (r= 0.572) and web visits (r= 0.570) should be noted.

Moreover, both electronic book and electronic serial expenditures unexpectedly correlate very low with web data. Indeed, web visits correlate better with book expenditures (r= 0.444) than with electronic book expenditures (r= 0.169).

If the analysis is repeated, taking only library websites into consideration (avoiding repositories, digital collections, and catalogs, 85 universities), the results are quite similar (Table 11), highlighting the good correlation of total expenditures with web data, and URL (G) and web visits with economic data.

Table 11. Correlation among institutional and web data (library websites; n = 85)

|  | Page count (Y) | PDF (G) | PDF (Y) | URL (G) | URL (Y) | LINK (M) | LINK (O) | Web visits (C) |
|---|---|---|---|---|---|---|---|---|
| Total expenditures | **0.509**\*\* | **0.340**\* | **0.270**\* | **0.546**\*\* | **0.381**\*\* | 0.177 | **0.215**\* | **0.573**\*\* |
| e-Book expenditures | 0.085 | 0.092 | 0.075 | 0.099 | 0.097 | 0.051 | 0.091 | 0.182 |
| Book expenditures | **0.284**\* | **0.255**\* | 0.158 | **0.417**\*\* | **0.367**\*\* | 0.084 | 0.166 | **0.450**\*\* |
| Serial expenditures | **0.334**\* | 0.213 | **0.226**\* | **0.351**\* | 0.200 | 0.174 | 0.094 | **0.390** |
| e-Serial expenditures | 0.206 | 0.160 | 0.147 | **0.248**\* | 0.061 | 0.065 | 0.107 | **0.266**\* |
| Physical visits | 0.138 | 0.147 | 0.084 | **0.245**\* | **0.298**\* | -0.031 | 0.103 | **0.286**\* |

\* Significant values (except diagonal) at the level of significance alpha=0.050 (two-tailed test)
\*\* Significant values (except diagonal) at the level of significance alpha=0.001 (two-tailed test)

**Discussion and conclusions**

The results obtained are discussed below. The main considerations covered are the following: the complexity of URL syntaxes of academic libraries, the problems of missing data for some indicators, the correlations obtained between web and economic data, and some technical limitations inherent to the web indicators used. Finally, general remarks are made, and good practices in the creation of URLs are proposed.

*Complexity of URL syntaxes*

A global analysis of the top 100 U.S. academic libraries according to their total expenditures has been performed around their web practices and performance. This analysis has been carried out at two different levels: institutional and web data.

The variety of University Library URL syntaxes (identified and classified in the method section) is wide, diverse and complex. Such practices produce a misrepresentation of academic libraries on the Web, and this diversity, coupled with other factors, may reduce the accuracy of web analysis.

Despite this deficiency, and taking into consideration the limitations exposed below, the library homepage is the most representative unit (43% of all considered units), followed by online catalogs (23%). However this result should be considered with some caution, because repositories, digital collections, and catalogs have been shown to be embedded into the library web-domain (without an independent sub-domain).

*Missing data*

Though the high performance of library homepages in the rankings by page count and visibility is clear, some differences are detected in page count (better performance of academic libraries) and visibility (better performance of digital collections and online catalogs).

This higher performance is expected both for libraries (suited to keep large amounts of data), and for digital collections (prone to be linked as quality reference sources), but is unexpected for online catalogs with complex URLs.

Apart from the limitation caused by complex URLs, as seen previously, another misrepresentation is detected due to the high percentage of URLs with missing data in some indicators, especially PDF page count (42% and 38% for Yahoo and Google respectively), and visits (44.4% for Compete). This is discussed in more detail below.

*Page count*

This effect could be explained by the fact that a great number of URLs tend not to contain PDF files:
- Most digital collections are composed of graphic and multimedia files.
- Most OPACs only show reference lists and do not link them to full texts (most of them in PDF format), due to legal and copyright issues of the library.
- Most academic library websites only publish online administrative content in html format.
- With respect to repositories, although these platforms are suited to store great amounts of PDF files, their performance in page count indicators is not as expected (except for some specific repositories, such as CiteseerX). One reason could be that only institutional repositories under sub-domains apart from the library are considered under this category. Other possible problems are the inability of search engine crawlers to collect data due to barriers in the design of the web databases and the use of other preferred formats beyond PDFs (Aguillo et. al 2010), and the use of non recommended metadata schemas (Arlitsch & O'Brian 2012)**.**

*Visits*

As regards total web visits, the following considerations could be set out:
- The OPAC URL in most cases is invisible for the user, and only appears after the query is submitted on the online platform or when results are displayed.
- The visits rate is low and the sources utilized do not have enough data to show.

*Web and institutional indicators*

With regard to institutional indicators, these show a logical correlation between total expenditures and the expenditures in printed material, both for books (r= 0.710) and serials (r= 0.775). Electronic material correlates worse with total expenditures, especially e-Books (r= 0.350).

Unexpectedly, the correlation among library expenditures and physical visits -although statistically significant- is low (r= 0.389): the libraries making a greater economical effort are not those which obtain more physical visits. Indeed, physical visits correlate very poorly among all other institutional indicators, especially with e-Books (r= 0.045).

The compactness among all web data indicators is also unexpectedly low. On one hand, page count correlates highly with links both for Google (r= 0.836) and Yahoo (r= 0.742). This result is consistent with those obtained by Tang and Thelwall (2008) who found that -for U.S. academic websites-, size (page count) was proportional to the number of inlinks. But on the other hand, link data (Majestic and Open Site Explorer) do not correlate with URL mention data (Google and Yahoo). This result is not consistent with previous works, which identify a good correlation among links and URL mentions (see Thelwall & Sud 2011). Although in these studies the link source was Yahoo, the conclusions were supposed to find a good correlation between indicators regardless of the source employed. For that reason, the poor visibility obtained is attributed to the complex structure and URL syntaxes identified previously, so that links are underrepresented due to the inability of web sources to retrieve all existing links at internal library web units. The good correlation between Link (M) and Link (O) (r= 0.823) reinforces this assumption.

Finally, if institutional and web data are compared, the correlations found are moderate. The best results are those that correlate total expenditures of the library with page count (r= 0.478), URL mentions measured by Google (r= 0.572), and visits (r= 0.570). These correlations are expected to be higher if URL structures and syntaxes are improved. Moreover, printed material correlates better with web data than electronic material (both for books and serials). The access to electronic documents via private platforms and the use of mobile devices may explain this result.

Notwithstanding, the correlations obtained are limited to the following considerations:
- The lack of an appropriate web policy within academic libraries, particularly related to the creation and design of sub-domains, limits the webometric analysis in several sources.
- Some of the web sources used are limited (such as the *Open site explorer* at the library level).
- Temporary differences exist between institutional and web data. In any case, expenditures and investments need some time to be reflected on the Web. In this sense, a periodic analysis –as performed in this research– will avoid this limitation by creating an appropriate statistical trend.
- The 100 U.S. universities with the highest total expenditures in academic libraries have been analyzed. It is possible that universities with lower expenditures in academic libraries may present different web performance so that a further analysis covering more universities should be carried out to check the results obtained.
- Web indicators used had technical limitations that should be considered. These are commented on below.

*Technical limitations*
Both sources (search engines) and web indicators (mainly URL mentions and links) present technical limitations that should be pointed out to contextualize results.

On one hand, search engines have been analyzed in-depth as webometric tools, and the following limitations have been found:

- Unpredictable fluctuation of hit counts, according to Rousseau (1999), Bar-Ilan (1999; 2001), and Lewandowski, Wahlig and Meyer-Bautor (2006).
- Lack of stability according to different types of queries (Uyar 2009a; 2009b).
- Differences in the hit count estimates obtained in the different search engine result pages (SERPs), partially attributed to the existence of duplicated or pseudo-duplicated content which is automatically eliminated (Gomes & Smith 2003; Thelwall 2008).

On the other hand, the web indicators used show limitations previously studied in the literature. The URL mentions are prone to show atypical points due to the fact that some short domain names may be included in other longer URLs (especially in e-mail addresses) when these data are extracted (Thelwall & Sud 2011; Thelwall, Sud & Wilkinson 2012). The page count and PDF count are sensitive to search engine fluctuations (and coverage), and web visits show limitations in the correct identification of different users and sessions (Bermejo 2007). All these limitations should be taken into account to properly interpret and contextualize the raw data obtained by the search engines.

*Final remarks*

Taking into account the aforementioned limitations, and despite the fact that the correlation values obtained are not highly significant (except for some specific indicators), the results are promising. Total expenditures seem to be a key indicator, but the correlations found are still weak. With regard to web data, great difficulties have been encountered in the gathering and measuring of data. If one is able to avoid linkage problems (due to the complexity of URLs), and to gain access to log files (for more accurate data about visits), these correlations could improve, which would be a first step in verifying the existence of a relationship between economic and web variables.

These web measurement difficulties must be understood at external access level (by means of the commercial search engines used). At internal level, webmasters may have tools to monitor both links and visits in a more accurate way.

Even so, if URLs are not built using logical structures (reflecting institutional hierarchies and using sub-domains for independent services and institutions) and syntaxes (avoiding non-friendly dynamic URLs), internal analyses are also jeopardized. If services and institutions under the umbrella of the academic library website are not assigned to independent URLs (sub-domains under the academic library), it is not possible to measure the web performance of each one independently.

For that reason, a hierarchical structure of URLs will allow an optimal external measurement of services, institutions, library branches and any other academic library units, necessary to check the impact of library services to web users and to carry out comparisons with other academic libraries.

Finally, the following good practices (partially aligned with those published by the Cybermetrics Lab)[2] are proposed in order to improve the web performance of academic libraries and their internal units and services:
- The Library should have a sub-domain within the university:
  <library.university.edu>

- Each unit, service, branch (including digital collections, and so on) should be included in a sub-domain under the library home page:
  <collection.library.university.edu>
- If the library unit considered becomes of general interest for the whole university, it should be created under a sub-domain independent of the Library (this is the case for institutional repositories:
  <repository.university.edu>

These URLs meet its two main tasks: access and identification. This hierarchy allows search engines to apply the same web indicator to all levels (unit, library and the whole university).

In any case, further research is needed in order to test new search engines, to expand both web and economic data, and to analyze more universities (both within U.S. and other countries), in order to compare and better contextualize the results obtained in this research.

**Notes**

[1] http://www.arl.org/bm~doc/code-of-best-practices-fair-use.pdf Accessed 26 February 2013.
[2] http://repositories.webometrics.info Accessed 26 February 2013.
[3] http://www.majesticseo.com Accessed 26 February 2013.
[4] http://www.opensiteexplorer.org Accessed 26 February 2013.
[5] http://ahrefs.com Accessed 26 February 2013.
[6] http://www.compete.com Accessed 26 February 2013.
[7] http://www.alexa.com Accessed 26 February 2013.
[8] http://hdl.handle.net/10481/23758 Accessed 26 February 2013.